\documentclass[twocolumn,preprintnumbers,amsmath,amssymb,superscriptaddress]{revtex4}
\usepackage{graphicx}
\usepackage{dcolumn}
\usepackage{bm}
\usepackage{soul}
\usepackage{color}
\usepackage{epstopdf}
\usepackage[version=3]{mhchem}
\usepackage{lipsum}
\usepackage[outercaption]{sidecap}
\usepackage{floatrow}
\usepackage{hyperref}
\usepackage{multirow}

\begin{document}

\title{Predicting Structural Relaxation in Supercooled Small Molecules via Molecular Dynamics Simulations and Microscopic Theory}
\author{Anh D. Phan}
\email{anh.phanduc@phenikaa-uni.edu.vn}
\affiliation{Faculty of Materials Science and Engineering, Phenikaa University, Hanoi 12116, Vietnam}
\affiliation{Phenikaa Institute for Advanced Study, Phenikaa University, Hanoi 12116, Vietnam}
\author{Ngo T. Que}
\affiliation{Phenikaa Institute for Advanced Study, Phenikaa University, Hanoi 12116, Vietnam}
\author{Nguyen T. T. Duyen}
\affiliation{Faculty of Materials Science and Engineering, Phenikaa University, Hanoi 12116, Vietnam}
\date{\today}

\begin{abstract}
Understanding and predicting the glassy dynamics of small organic molecules is critical for applications ranging from pharmaceuticals to energy and food preservation. In this work, we present a theoretical framework that combines molecular dynamics simulations and Elastically Collective Nonlinear Langevin Equation (ECNLE) theory to predict the structural relaxation behavior of small organic glass-formers. By using propanol, glucose, fructose, and trehalose as model systems, we estimate the glass transition temperature ($T_g$) from stepwise cooling simulations and volume–temperature analysis. These computed $T_g$ values are then inserted into the ECNLE theory to calculate temperature-dependent relaxation times and diffusion coefficients. Numerical results agree well with experimental data in prior works. This approach provides a predictive and experimentally-independent route for characterizing glassy dynamics in molecular materials.
\end{abstract}

\maketitle
\section{Introduction}
Small organic molecules such as glucose, fructose, trehalose, and propanol have been extensively studied due to their biological relevance and wide-ranging applications in food science, biotechnology, and pharmaceuticals \cite{intro1, intro2, intro3, intro4}. Glucose and fructose are monosaccharides that serve as key energy sources and metabolic intermediates in living organisms \cite{intro5}. Fructose, which is naturally present in fruits and honey, is notably sweeter than glucose and is frequently used as a sweetening agent. Both sugars are commonly incorporated into products such as soft drinks, confectionery, and dairy-based foods \cite{intro3}. Trehalose, a disaccharide composed of two glucose units, is recognized for its unique ability to stabilize proteins and biological membranes under stress conditions. This protective functionality has led to its increasing use in food preservation and pharmaceutical formulations \cite{intro1, intro2}. Propanol, an aliphatic alcohol, also plays a functional role in the food industry, where it is produced during fermentation and contributes to flavor development and product stability\cite{intro4}.

The glass transition temperature is a fundamental thermophysical property of amorphous materials that indicates the transition between the rigid, glassy state and the more mobile, rubbery state. It is highly sensitive to molecular characteristics such as chain stiffness, molecular weight, and intermolecular interactions. Importantly, $T_g$ is intimately linked to the structural relaxation time, $\tau_\alpha(T)$, which characterizes the timescale of molecular rearrangements. As the temperature decreases below $T_g$, $\tau_\alpha$ increases dramatically and this corresponds to the slowing down of molecular motion and the emergence of kinetic arrest characteristic of glass-forming systems. Understanding this relationship is essential for predicting temperature-dependent dynamics and stability in amorphous molecular materials.

Although experimental methods including differential scanning calorimetry, dynamic mechanical analysis, and broadband dielectric spectroscopy (BDS) are commonly used to measure $T_g$, each has inherent limitations. The DSC is widely accessible and can measure $T_g$ under varying cooling rates and pressures \cite{DSC}, but its accuracy may be limited by overlapping thermal events such as melting which can obscure the glass transition signal. In addition, this method cannot measure the temperature dependence of the structural relaxation time. The dynamic mechanical analysis probes the viscoelastic response of materials to capture changes in storage and loss moduli as a function of temperature. However, it requires careful sample preparation to ensure accurate mechanical contact and reproducible measurements \cite{DMA}. The broadband dielectric spectroscopy provides detailed insight into molecular relaxation dynamics by measuring the frequency-dependent dielectric response, which directly describe dipolar reorientations and segmental motions \cite{BDS}. Despite its ability to probe molecular dynamics over a broad frequency and temperature range, BDS is less commonly employed due to its specialized instrumentation requirements and the need for materials with sufficient dielectric contrast. These limitations have led to growing interest in molecular dynamics (MD) simulations and statistical mechanical theories as molecularly detailed and scalable tools for studying glass transition behavior \cite{ECNLE1, ECNLE2, ECNLE16}. While MD simulations provide valuable insight into atomistic mechanisms, it is inherently restricted by its limited timescale, typically reaching only up to $10^5$ picoseconds \cite{timescale}. As a result, it cannot directly capture the slow relaxation dynamics that are commonly observed in experiments near and below the glass transition temperature. Bridging this temporal gap remains a key challenge: how can we extend the reach of MD-based approaches to accurately predict the long-timescale dynamics characteristic of experimental observations in glass-forming systems?

To overcome the timescale limitations of MD simulations, we employ the ECNLE theory which is a statistical mechanical framework capable of predicting structural relaxation times and diffusion coefficients across a broad temporal range from picoseconds to hundreds of seconds \cite{ECNLE1, ECNLE2, ECNLE3, ECNLE4, ECNLE5, ECNLE6, ECNLE7, ECNLE8, ECNLE9, ECNLE10, ECNLE11, ECNLE12, ECNLE13, ECNLE14, ECNLE15, ECNLE16, ECNLE17, ECNLE18}. The ECNLE theory combines local cage-scale dynamics with long-range collective elastic effects to describe thermally activated relaxation in glass-forming systems. It requires only minimal input parameters relying primarily on the equilibrium static structure and a thermal mapping governed by the glass transition temperature. The theory has been successfully applied to describe the temperature and pressure dependence of relaxation dynamics in a wide variety of amorphous materials, including polymers \cite{ECNLE2, ECNLE3, ECNLE4, ECNLE5, ECNLE10, ECNLE11}, thermal liquids \cite{ECNLE9}, metallic glasses \cite{ECNLE15, ECNLE16, ECNLE18}, and organic materials  \cite{ECNLE2, ECNLE3, ECNLE4, ECNLE5, ECNLE11} with strong quantitative agreement with both simulations and experimental measurements. However, one persistent challenge is that experimental $T_g$ values are not always available, particularly for newly developed or computationally designed materials. This limits the applicability of such models in the absence of empirical calibration. 

In recent work \cite{ECNLE16}, we have employed MD simulations to compute values for metallic glasses and incorporated them into the theoretical framework of structural relaxation. The resulting predictions for $\tau_\alpha(T)$ obtained from the integrated MD–theory approach showed excellent quantitative agreement with simulation data. This success raises several important questions. Since the accessible timescale in MD simulations is typically limited as shown in Fig. \ref{fig1}, can the same integrated MD–theory approach accurately describe relaxation dynamics over the much longer timescales probed in experiments ranging from milliseconds to hundreds of seconds? More specifically, does the theory retain its predictive ability when benchmarked against experimental data, where relaxation times span many orders of magnitude beyond what simulations can directly access? Can the same integrated approach be extended to organic molecular systems?

In this work, we address these questions by combining MD simulations and ECNLE theory to investigate the glassy dynamics of four representative small-molecule glass formers: propanol, glucose, fructose, and trehalose. We first compute the glass transition temperature for each compound via stepwise MD cooling simulations and volume–temperature analysis. These simulation-derived $T_g$  values and their experimental counterparts are then used as inputs to ECNLE theory to predict the temperature dependence of structural relaxation times and diffusion coefficients. The theoretical predictions are evaluated against available experimental data to assess their accuracy and applicability. Through this integrated framework, we extend the predictive reach of molecular simulations, provide new insights into the relaxation behavior of organic molecular glasses, and establish a general computational strategy for the design and screening of amorphous materials.

\section{Theoretical background of the ECNLE theory}

\begin{figure*}[htp]
\includegraphics[width=18cm]{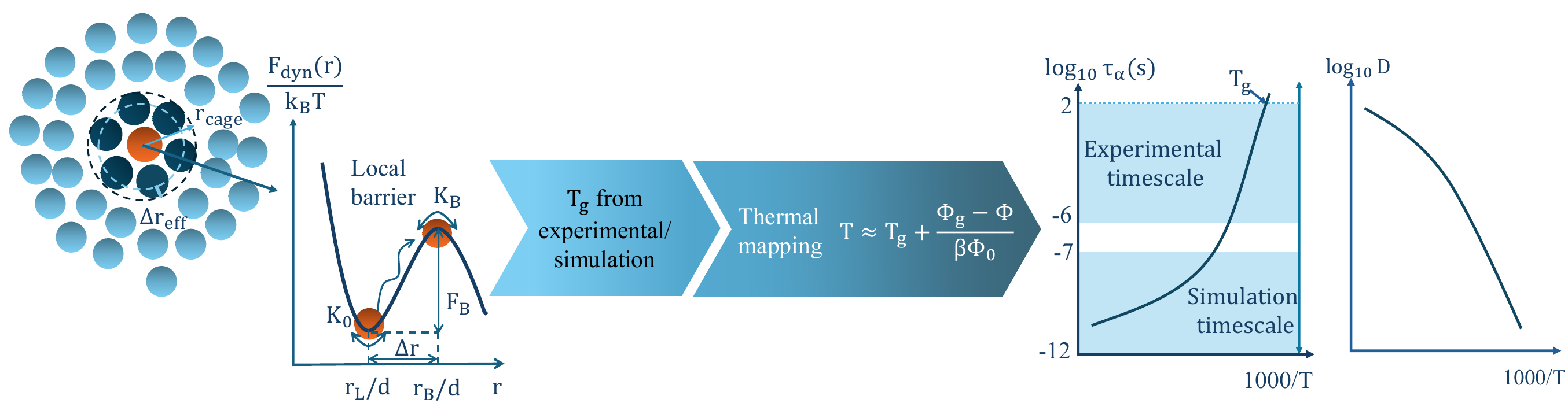}
\caption{\label{fig1}(Color online) Schematic illustration of the dynamic free energy in the ECNLE theory. Key length and energy quantities are defined.}
\end{figure*}
In the ECNLE theory, amorphous materials are theoretically modeled as hard-sphere liquids to capture essential features of their local structure and dynamic caging behavior \cite{ECNLE1, ECNLE2, ECNLE3, ECNLE4, ECNLE5, ECNLE6, ECNLE7, ECNLE8, ECNLE9, ECNLE10, ECNLE11, ECNLE12, ECNLE13, ECNLE14, ECNLE15, ECNLE16, ECNLE17, ECNLE18,ECNLE19,ECNLE20,ECNLE21,Xie}. The hard-sphere liquid is characterized by the particle number density, $\rho$, the particle diameter, $d$, and the volume fraction $\Phi = \pi \rho d^3/6$. The static structure factor, $S(q)$, and the radial distribution function, $g(r)$, are computed using the Percus–Yevick approximation \cite{Percus-Yevick}. The dynamics of a tagged particle is primarily governed by interactions with its surrounding neighbors, stochastic thermal forces, and friction \cite{ECNLE1, ECNLE2, ECNLE3, ECNLE4, ECNLE5, ECNLE6, ECNLE7, ECNLE8, ECNLE9, ECNLE10, ECNLE11, ECNLE12, ECNLE13, ECNLE14, ECNLE15, ECNLE16, ECNLE17, ECNLE18,ECNLE19,ECNLE20,ECNLE21,Xie}. These influences are described by a nonlinear stochastic equation of motion. Solving this equation yields the dynamic free energy, which quantifies the effective interaction between the tagged particle and its nearest neighbors and governs its local mobility \cite{ECNLE1, ECNLE2, ECNLE3, ECNLE4, ECNLE5, ECNLE6, ECNLE7, ECNLE8, ECNLE9, ECNLE10, ECNLE11, ECNLE12, ECNLE13, ECNLE14, ECNLE15, ECNLE16, ECNLE17, ECNLE18,ECNLE19,ECNLE20,ECNLE21,Xie}.
\begin{eqnarray} 
\frac{F_{\text{dyn}}(r)}{k_B T} &=& -3\ln \left(\frac{r}{d}\right) \\
&-& \int_0^\infty \frac{q^2 d^3 [S(q)-1]^2}{12\pi \Phi [1 + S(q)]} \exp\left[-\frac{q^2 r^2 (1 + S(q))}{6 S(q)}\right] dq,\nonumber
\label{eq:1}
\end{eqnarray}
where $k_B$ is the Boltzmann constant, $T$ is the ambient temperature, $r$ is the displacement, and $q$ is the wavevector. The second term in the dynamic free energy captures the caging constraint imposed by neighboring particles, while the first term accounts for properties of an ideal fluid.

The dynamic free energy provides several key physical quantities that determine the local dynamics of a particle including the barrier height, localization length, and barrier position. At low volume fractions ($\Phi < 0.432$), interactions between a tagged particle and its nearest neighbors are sufficiently weak to allow particles to move freely, and the system behaves as a fluid \cite{ECNLE1, ECNLE2, ECNLE3, ECNLE4, ECNLE5, ECNLE6, ECNLE7, ECNLE8, ECNLE9, ECNLE10, ECNLE11, ECNLE12, ECNLE13, ECNLE14, ECNLE15, ECNLE16, ECNLE17}. As the volume fraction increases ($\Phi \geq 0.432$), reduced interparticle separation gives rise to transient cage-like configurations that dynamically confine each particle. The radius of this cage, $r_{\text{cage}}$, is determined by the position of the first local minimum in the radial distribution function $g(r)$. This confinement leads to the emergence of a local energy barrier in the dynamic free energy as illustrated in Fig.~\ref{fig1}. The height of this barrier is quantified by $F_B = F_{\text{dyn}}(r_B) - F_{\text{dyn}}(r_L)$, where $r_L$ is the localization length and $r_B$ is the barrier position. 

During the escape of a tagged particle from its surrounding cage, the particles in the first coordination shell must undergo cooperative rearrangement. This collective motion induces a displacement field, $u(r)$, which originates at the particle surface and propagates radially outward from the cage boundary into the surrounding medium. According to Lifshitz’s continuum elasticity theory \cite{Lifshitz}, this displacement field for distances $r \geq r_{cage}$ can be described analytically as
\begin{eqnarray} 
u(r) = \Delta r_{eff}\frac{r_{cage}^2}{r^2},
\label{eq:2}
\end{eqnarray} 
where $\Delta r_{eff}$ is the effective amplitude of the displacement field calculated using \cite{ECNLE7, ECNLE9}
\begin{eqnarray} 
\Delta r_{eff} = \frac{3}{r_{cage}^3}\left[\frac{r_{cage}^2\Delta r^2}{32}- \frac{r_{cage}\Delta r^3}{192}+\frac{\Delta r^4}{3072}\right],
\label{eq:3}
\end{eqnarray}
where $\Delta r = r_B - r_L$ is the jump distance between the barrier location and the localization length.

Given that the displacement field $u(r)$ is typically small, the particle motion can be approximated as harmonic oscillations around their equilibrium positions. This assumption allows the local elastic energy associated with a small displacement to be expressed as $\cfrac{1}{2}K_0 u^2(r)$, where $K_0 = \left. \cfrac{\partial^2 F_{\text{dyn}}(r)}{\partial r^2} \right|_{r = r_L}$ is the effective spring constant. This local spring constant characterizes the stiffness of the confining cage and quantifies the resistance of a particle to displacement from its localized state. In the ECNLE theory, the structural relaxation involves not only the local barrier associated with cage escape but also a long-range collective elastic distortion of the surrounding medium. These small but extended displacements induce an additional barrier known as the elastic barrier. This cooperative barrier is computed by integrating the local elastic energy over all particles outside the cage region \cite{ECNLE1, ECNLE2, ECNLE3, ECNLE4, ECNLE5, ECNLE6, ECNLE7, ECNLE8, ECNLE9, ECNLE10, ECNLE11, ECNLE12, ECNLE13, ECNLE14, ECNLE15, ECNLE16, ECNLE17, ECNLE18}:
\begin{eqnarray}
F_e = 4\pi \rho \int_{r_{\text{cage}}}^{\infty} r^2 g(r) \frac{K_0 u^2(r)}{2} \, dr.
\label{eq:4}
\end{eqnarray} 
Equation (\ref{eq:4}) captures the coupling between the local dynamics and collective motions and provides the key to understanding the dramatic slowdown of dynamics near the glass transition. By using Kramers’ theory, we can quantitatively compute the structural relaxation time $\tau_{\alpha}$ in supercooled liquids and glasses via \cite{ECNLE1, ECNLE2, ECNLE3, ECNLE4, ECNLE5, ECNLE6, ECNLE7, ECNLE8, ECNLE9, ECNLE10, ECNLE11, ECNLE12, ECNLE13, ECNLE14, ECNLE15, ECNLE16, ECNLE17, ECNLE18,ECNLE19,Xie}
\begin{eqnarray} 
\frac{\tau_\alpha}{\tau_s} = 1 + \frac{2\pi}{\sqrt{K_0 K_B}} \frac{k_B T}{d^2} \exp\left(\frac{F_B + F_e}{k_B T}\right), 
\label{eq:5}
\end{eqnarray}
where $K_B = \left| \cfrac{\partial^2 F_{\text{dyn}}(r)}{\partial r^2} \right|_{r = r_B}$ is the curvature of the dynamic free energy at the barrier position and $\tau_s$ is a short relaxation timescale. The analytical form of $\tau_s$ is \cite{ECNLE1, ECNLE2, ECNLE3, ECNLE4, ECNLE5, ECNLE6, ECNLE7, ECNLE8, ECNLE9, ECNLE10, ECNLE11, ECNLE12, ECNLE13, ECNLE14, ECNLE15, ECNLE16, ECNLE17, ECNLE18,ECNLE19,Xie}
\begin{eqnarray} 
\tau_s = g^2(d)\tau_E\left[1+\frac{1}{36\pi\Phi}\int_0^\infty dq\frac{q^2(S(q)-1)^2}{S(q)+b(q)} \right],
\label{eq:6}
\end{eqnarray}
where $1/b(q) = 1 - j_0(q) + 2j_2(q)$ with $j_n(x)$ being the spherical Bessel function of order $n$, $g(d)$ is the pair correlation function at contact, and $\tau_E$ is the Enskog time scale expressed by \cite{ECNLE8,ECNLE10}
\begin{eqnarray} 
\tau_E = \frac{d}{24\Phi g(d)}\sqrt{\frac{\pi M}{k_BT}},
\label{eq:6-1}
\end{eqnarray}
where $M$ is the molecular weight of the material. The effective molecular diameter $d$ was determined by generating three-dimensional molecular structures from the Simplified Molecular Input Line Entry System (SMILES) strings \cite{smiles} using the RDKit package \cite{RDKit}, and then calculating the spherical-equivalent diameter from the molecular volume. The molecular parameters employed are glucose ($d = 7.42$~\AA, $M = 180.156$g/mol), fructose ($d = 6.41$\AA, $M = 180.156$g/mol), trehalose ($d = 9.34$\AA, $M = 342.297$g/mol), and propanol ($d = 4.29$\AA, $M = 60.096$~g/mol). As the temperature increases from $0.6T_g$ to $1.5T_g$, $\tau_E$ varies from approximately $0.5 \times 10^{-13}$~s to $2 \times 10^{-13}$~s. However, this variation has a negligible effect on the temperature dependence of structural relaxation time, particularly in the low-temperature regime relevant to experimentally accessible conditions (see Fig. S1 in Supplementary Materials). Therefore, to be consistent with previous studies on a wide range of amorphous materials including molecular liquids, polymers, metallic glasses, and amorphous pharmaceuticals \cite{ECNLE1, ECNLE2, ECNLE3, ECNLE4, ECNLE5, ECNLE6, ECNLE7, ECNLE8, ECNLE9, ECNLE10, ECNLE11, ECNLE12, ECNLE13, ECNLE14, ECNLE15, ECNLE16, ECNLE17, ECNLE18, ECNLE19, ECNLE20, ECNLE21}, we adopt a zeroth-order approximation of $\tau_E = 10^{-13}$~s in all calculations.

Note that Eqs. (\ref{eq:1}-\ref{eq:6}) allow us to determine the volume-fraction dependence of the structural relaxation, $\tau_\alpha(\Phi)$. To quantitatively compare with experimental data or simulation, we convert the volume fraction into temperature by using a thermal mapping, which is constructed from the thermal expansion process \cite{ECNLE16,ECNLE14,ECNLE15,ECNLE18,ECNLE19,ECNLE20,ECNLE21}
\begin{eqnarray} 
T \approx T_g + \frac{\Phi_g - \Phi}{\beta \Phi_0},
\label{eq:7}
\end{eqnarray}
where $T_g$ and $\Phi_g=0.6157$ are the glass transition temperature and the volume fraction at $\tau_\alpha=100s$, respectively, $\Phi_0=0.5$ is a characteristic volume fraction, $\beta \approx 12 \times 10^{-4}$ is an effective thermal expansion coefficient which is assumed to be approximately constant across a wide range of amorphous materials \cite{ECNLE16,ECNLE14,ECNLE15,ECNLE18,ECNLE19,ECNLE20,ECNLE21}. The glass transition temperature can be determined through experiments \cite{ECNLE14,ECNLE15,ECNLE18,ECNLE19,ECNLE21} or MD simulations \cite{ECNLE1, ECNLE2, ECNLE16}, or, more recently, machine/deep learning \cite{ECNLE18,ECNLE19}. Most prior studies have relied on the experimental $T_g$ to successfully describe BDS data and give good quantitative agreement across a wide range of glass-forming systems. For novel or unsynthesized materials, experimental $T_g$ values are frequently lacking. In the following sections, we show how MD simulations can be used to estimate $T_g$  for several organic glass formers and how these values can be utilized within the theoretical framework to predict temperature-dependent structural relaxation over experimentally relevant timescales.

Furthermore, the temperature dependence of the diffusion coefficient can be determined from the structural relaxation time and the jump distance using the following relation \cite{ECNLE16,ECNLE20,ECNLE21}
\begin{eqnarray}
D(T)=\frac{\Delta r^2}{6\tau_{\alpha} (T)}.
\label{eq:8}
\end{eqnarray}

\section {Molecular Dynamics Simulation}
To determine the glass transition temperature of glass-forming systems such as propanol, glucose, fructose, and trehalose, MD simulations were carried out using the Large-scale Atomic/Molecular Massively Parallel Simulator \cite{LAMMPS}. The simulation workflow comprised three essential stages: (1) initial structure generation and energy minimization, (2) thermalization and cooling via the NPT ensemble, and (3) $T_g$ estimation from the temperature-dependent volume profile. 

Initial structures of propanol and sugar molecules were constructed and placed in a cubic simulation box of dimensions  $60\times60\times60\AA^3$ under periodic boundary conditions in all directions. The simulation cell was then packed to achieve an initial density close to 1.0 $g/cm^3$. We employed the Dreiding force field since it has been widely used to investigate both organic molecules and polymeric systems \cite{force_field}. This force field includes bonded interactions (bond stretching, angle bending, and dihedral torsions) and nonbonded terms (van der Waals and hydrogen bonding). This comprehensive interaction provides sufficient flexibility to model the structural and thermal behavior of small organic molecules and sugars. Following energy minimization, each system was equilibrated in the NPT ensemble at a high initial temperature chosen to ensure sufficient molecular mobility and thermal equilibration. The temperatures used were 140 K for propanol, 350 K for both glucose and fructose, and 500 K for trehalose. Each system was equilibrated for 250 ps using the Nosé–Hoover thermostat and barostat to maintain constant temperature and pressure conditions. After reaching equilibrium, a stepwise cooling procedure was applied. The temperature was gradually reduced from the initial equilibration temperature to a final target temperature, which was set to 40 K for propanol, 200 K for glucose, 150 K for fructose, and 300 K for trehalose. The cooling process was carried out over 20 equally spaced temperature intervals to ensure sufficient resolution of the temperature-dependent volume behavior. At each temperature, the system was equilibrated for 200 ps, and the average specific volume was calculated from the final 100 ps. The resulting volume–temperature data were then used to determine $T_g$ by fitting two linear regimes corresponding to the high-temperature liquid-like state and the low-temperature glassy state to identify their intersection point.

\section{Results AND Discussion}
\begin{figure*}[htp]
\includegraphics[width=16cm]{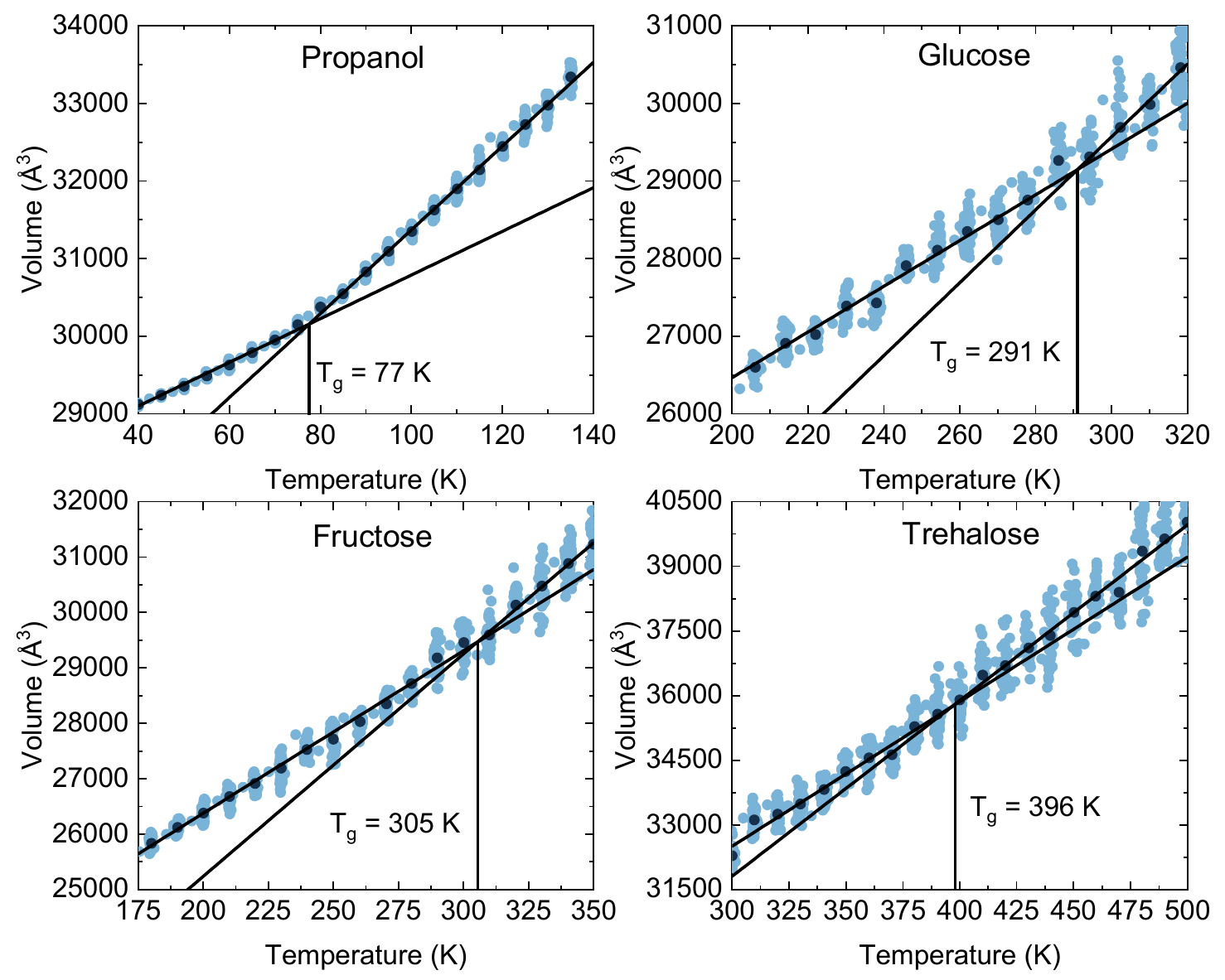}
\caption{\label{fig2}(Color online) Temperature dependence of the volume of the simulation box obtained from MD simulations for (a) propanol, (b) glucose, (c) fructose, and (d) trehalose. Blue circles indicate raw simulation data collected at each temperature step, while black circles are averaged volume values used for analysis. Linear fits to the high-temperature (liquid-like) and low-temperature (glassy) regimes are shown as solid lines. The intersection of these lines defines the estimated glass transition temperature for each material.}
\end{figure*}

Figure \ref{fig2} shows the temperature dependence of the volume of our simulation box obtained from MD simulations for the four studied systems. As the temperature decreases, each system exhibits a characteristic change in the slope of the volume–temperature curve, which reflects the transition from a supercooled liquid state with relatively high thermal expansivity to a glassy state with significantly slowed-down molecular mobility. By performing linear fits to the averaged volume data in the high-temperature and low-temperature regimes, the intersection point of these lines gives the glass transition temperature for each material. The resulting $T_g$ values are 77 K for propanol, 291 K for glucose, 305 K for fructose, and 396 K for trehalose. Table \ref{table1} compares our MD-predicted $T_g$ values with experimental data from previous studies \cite{Tg_propanol_96.2K,Tg_propanol_98K, Tg_Glucose_295K_Trehalose_368K, Tg_glucose_312K, Fructose_280K, Fructose_286K, Trehalose_379K}. The close agreement across all four materials validates our simulation and force field parameters in capturing the thermophysical behavior of amorphous molecular materials. These computational $T_g$ values can now be directly used as input into the thermal mapping equation of the ECNLE theory (Eq.~(\ref{eq:7})) to predict the temperature dependence of the structural relaxation time without requiring experimental input.
  
\begin{table}[hpt]
    \centering
    \caption{\label{table1}The glass transition temperatures in Kelvin computed in our work and their corresponding experimental values in previous works.}
    \begin{tabular}{|c|c|c|}
    \hline
        Material & $T_g$ (this work) & $T_g$ (prior work)  \\ \hline
        Propanol & 77 & 96.2 \cite{Tg_propanol_96.2K}, 98 \cite{Tg_propanol_98K}, 102\cite{relaxation_time_propanol}  \\ \hline
        Glucose & 291 & 295 \cite{Tg_Glucose_295K_Trehalose_368K}, 312 \cite{Tg_glucose_312K}, 308\cite{relaxation_time_glucose_fructose}  \\ \hline
        Fructose & 305 & 280 \cite{Fructose_280K}, 286 \cite{Fructose_286K}, 288\cite{relaxation_time_glucose_fructose} \\ \hline
        Trehalose & 396 & 368 \cite{Tg_Glucose_295K_Trehalose_368K}, 379 \cite{Trehalose_379K}, 371\cite{relaxation_time_trehalose}  \\ \hline
    \end{tabular}
\end{table}

\begin{figure*}[htp]
\includegraphics[width=16cm]{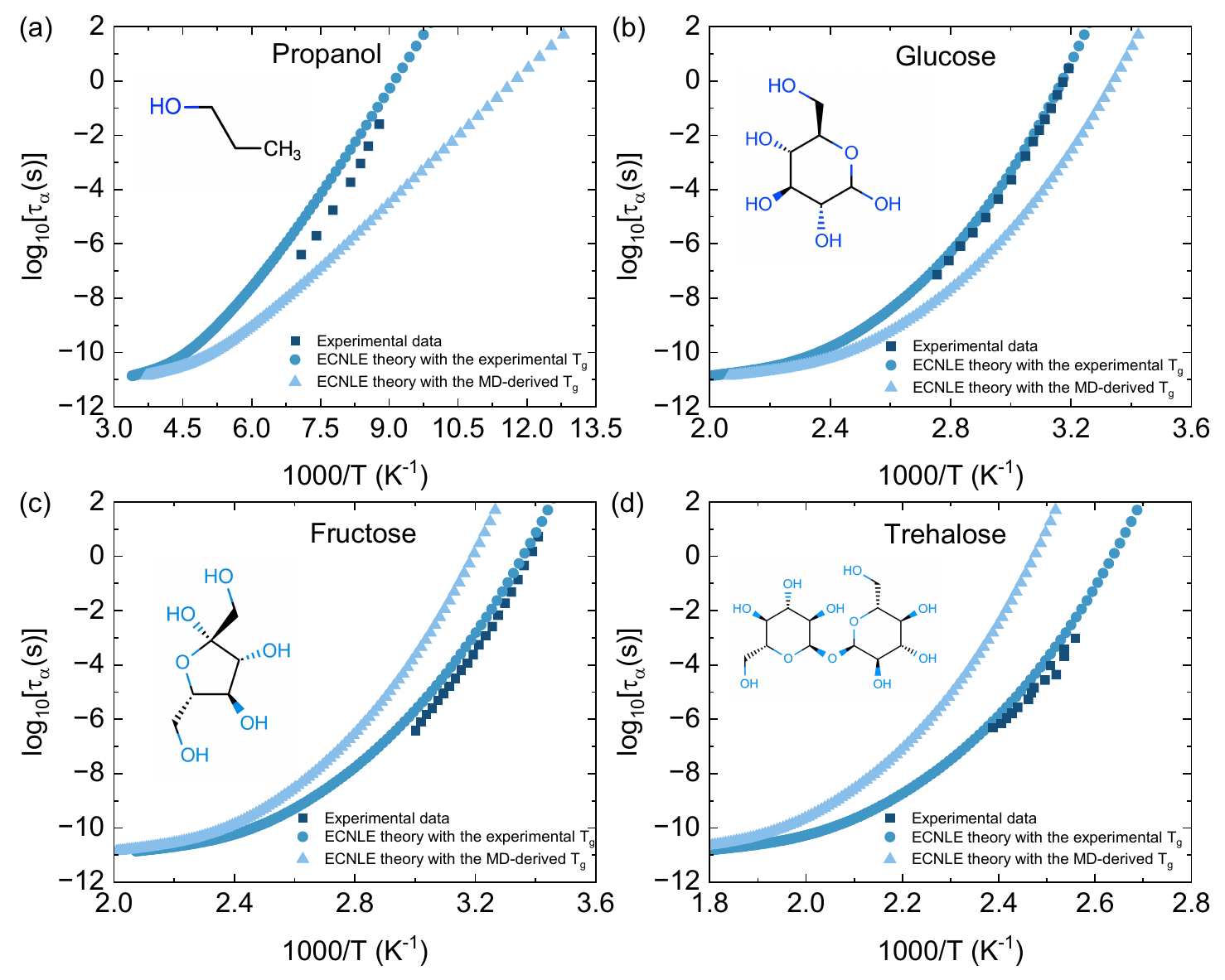}
\caption{\label{fig3}(Color online) Logarithm of the structural relaxation time as a function of $1000/T$ for (a) propanol, (b) glucose, (c) fructose, and (d) trehalose. Square symbols are experimental $\tau_\alpha(T)$ data taken from Refs.~\cite{relaxation_time_propanol, relaxation_time_glucose_fructose, relaxation_time_trehalose}. Circle and triangle symbols correspond to the ECNLE predictions using experimental $T_g$s in Refs.~\cite{relaxation_time_propanol, relaxation_time_glucose_fructose, relaxation_time_trehalose} and the $T_g$ obtained from MD simulations in this work, respectively.}
\end{figure*}

Figure \ref{fig3} presents the temperature dependence of the structural relaxation time for propanol, glucose, fructose, and trehalose. We calculate the relaxation time as a function of the volume fraction, $\tau_\alpha(\Phi)$, by using Eqs. (\ref{eq:1})–(\ref{eq:6}). To enable direct comparison with experimental data, the volume fraction is then converted to temperature using the thermal mapping relation defined in Eq.(\ref{eq:7}), where the $T_g$ value is a critical input parameter. For each system, we use both the $T_g$ values obtained from our simulations and experiments in Refs.~\cite{relaxation_time_propanol, relaxation_time_glucose_fructose, relaxation_time_trehalose} as summarized in Table~\ref{table1}. Our ECNLE calculations are quantitatively compared with prior experimental studies {\cite{relaxation_time_propanol, relaxation_time_glucose_fructose, relaxation_time_trehalose}}. For propanol and glucose, the relaxation dynamics predicted using simulation-estimated $T_g$ values show good agreement with the experimental data over a full temperature range. In the case of fructose and trehalose, the predictions using experimental $T_g$ values more accurately capture the low-temperature behavior of $\tau_\alpha(T)$. Nevertheless, the ECNLE predictions based on simulation-estimated $T_g$ values remain quantitatively reasonable and capture the temperature dependence of relaxation dynamics, particularly when experimental measurements are unavailable. 

\begin{figure*}[h]
    \centering
    \includegraphics[width=16cm]{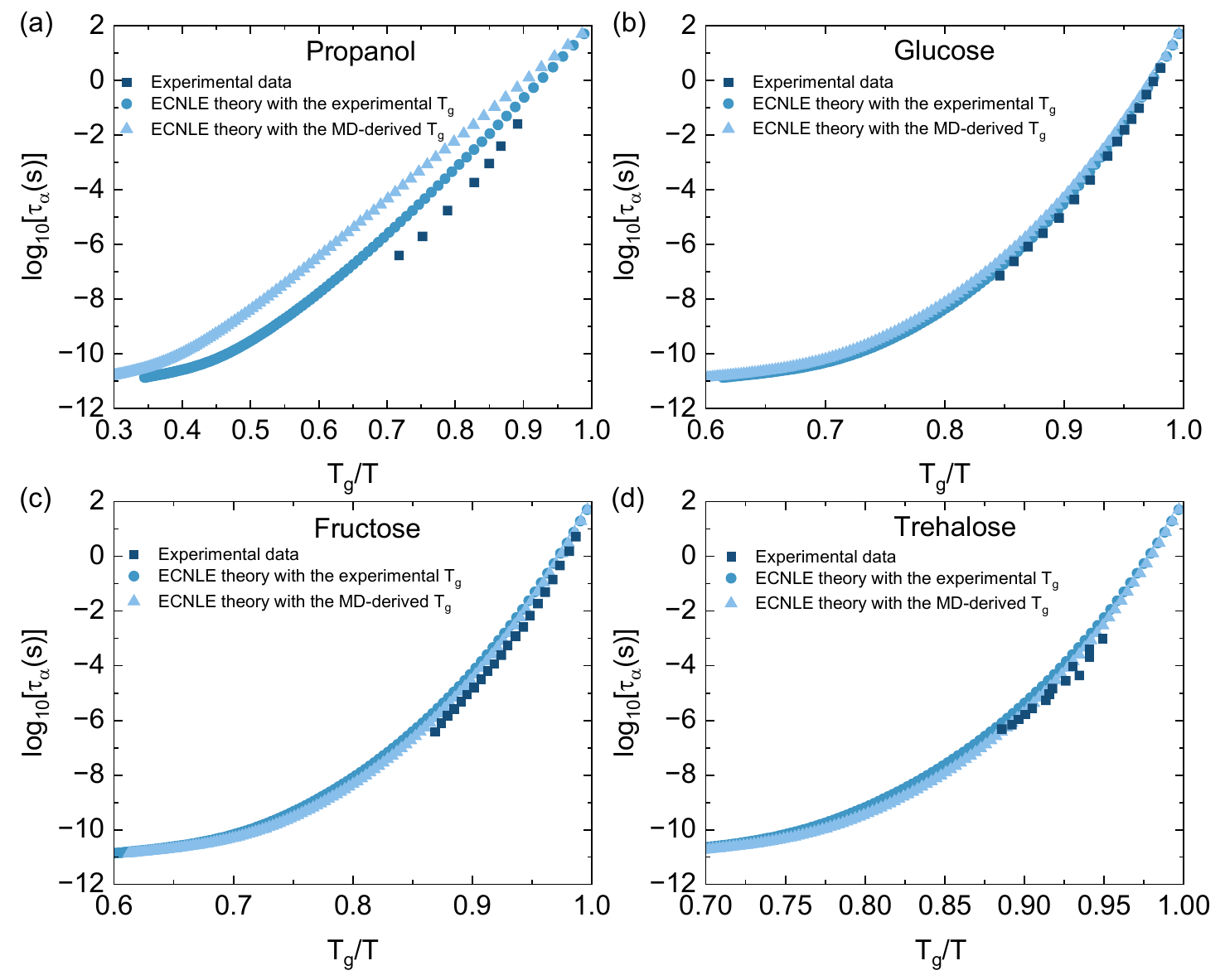} 
    \caption{(Color online) Logarithm of the structural relaxation time as a function of $T_g/T$ for (a) propanol, (b) glucose, (c) fructose, and (d) trehalose. Square symbols represent experimental data. Curves with circle and triangle symbols correspond to the ECNLE predictions using the $T_g$s from our MD simulation and experiments, respectively.}
    \label{fig:fig4} 
\end{figure*}

To further evaluate the reliability and predictive accuracy of our method, the data from Fig.~\ref{fig3} are replotted in the Angell representation in Fig.~\ref{fig:fig4}, where $\log_{10}[\tau_\alpha(T)]$ is plotted against the normalized inverse temperature $T_g/T$ to facilitate comparison of dynamic fragility. Although the experimental and simulation-derived $T_g$ values differ due to the high cooling rates and the limited sampling timescales and this causes noticeable deviations in $\tau_\alpha(T)$ when plotted versus $1000/T$, the Angell plots reveal that our ECNLE predictions are in good quantitative agreement with experimental relaxation times. Although direct simulation data for $\tau_\alpha(T)$ and $D(T)$ are not available for the small organic molecules examined in this study, our focus is to evaluate the ECNLE calculations using MD-derived $T_g$ values and compare them with experimental relaxation and diffusion data without any adjustable parameter. Previous ECNLE studies \cite{ECNLE16, ECNLE20} have already shown quantitative agreement with MD simulations for many organic materials and metallic glasses. Thereby, this indicates that the ECNLE theory, when combined with MD-estimated $T_g$ values, can reliably describe the relaxation dynamics and fragility behavior of glass-forming systems.

\begin{figure*}[h]
\centering
\includegraphics[width=16cm]{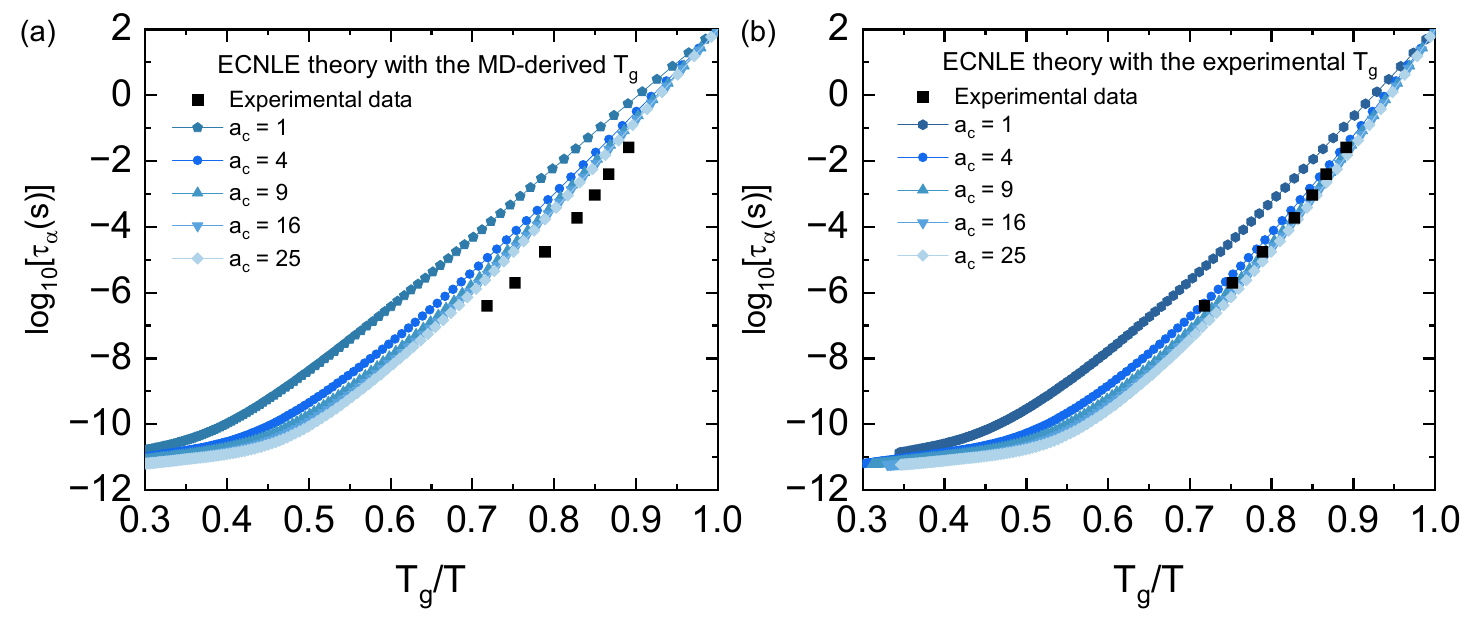}
\caption{\label{fig6}(Color online) {Logarithm of the structural relaxation time as a function of $T_g/T$ for propanol, calculated using Eqs. (\ref{eq:1})–(\ref{eq:6}), (\ref{eq:7}), and (\ref{eq:9}) with both simulation-derived (a) and experimental (b) $T_g$ values. Results are shown for different values of $a_c$. Experimental data are included for comparison.}}
\end{figure*}

As shown in Fig.~\ref{fig:fig4}, propanol exhibits a large difference in fragility with experimental data indicating a value around 25, whereas ECNLE calculations based on the MD-derived $T_g$ predict a slightly higher value of approximately 31. This discrepancy may arise from a fundamental limitation of our ECNLE approach, which assumes a universal relationship between the local barrier and the collective elastic barrier. It implies that the interplay between short-range and long-range dynamics is material-independent. However, such the assumption may not fully capture chemical and structural specificities that govern relaxation behavior in diverse glass-forming systems. To address this Xie and Schweizer \cite{Xie} proposed an extended ECNLE approach that the coupling between local hopping and elastic constraints is no longer universal but instead depends on molecular-scale features. They introduced a material-specific parameter, $a_c$, to scale the elastic barrier as $F_{e,\text{new}} = a_c F_e$. This modification adjusts the relative contribution of collective elasticity to the total barrier. Now, the temperature dependence of structural relaxation time is calculated by 
\begin{eqnarray} 
\frac{\tau_\alpha}{\tau_s} = 1 + \frac{2\pi}{\sqrt{K_0 K_B}} \frac{k_B T}{d^2} \exp\left(\frac{F_B + a_cF_e}{k_B T}\right). 
\label{eq:9}
\end{eqnarray}
Here, $a_c$ is manually tuned to obtain the best fit with experimental data. Figure \ref{fig6} presents the Angell plots of propanol, comparing experimental measurements with ECNLE predictions using both the MD-derived $T_g$ (Fig. \ref{fig6}a) and the experimental $T_g$ (Fig. \ref{fig6}b). As $a_c$ increases, the collective elastic contribution becomes more significant. It leads to a steeper temperature dependence of $\tau_\alpha$ and improves agreement with the experimental variation. In particular, when $a_c$ is around 9-25 and the experimental $T_g$ is used in the thermal mapping (Eq.~(\ref{eq:7})), the ECNLE curves perfectly overlap experimental data. This extension allows us to quantify the role of collective elasticity in governing the glass transition and fragility in chemically diverse systems. However, we have to emphasize that our present work aims to predict $\tau_\alpha(T)$ and the dynamic fragility without relying on empirical fitting.

\begin{figure*}[htp]
\includegraphics[width=16cm]{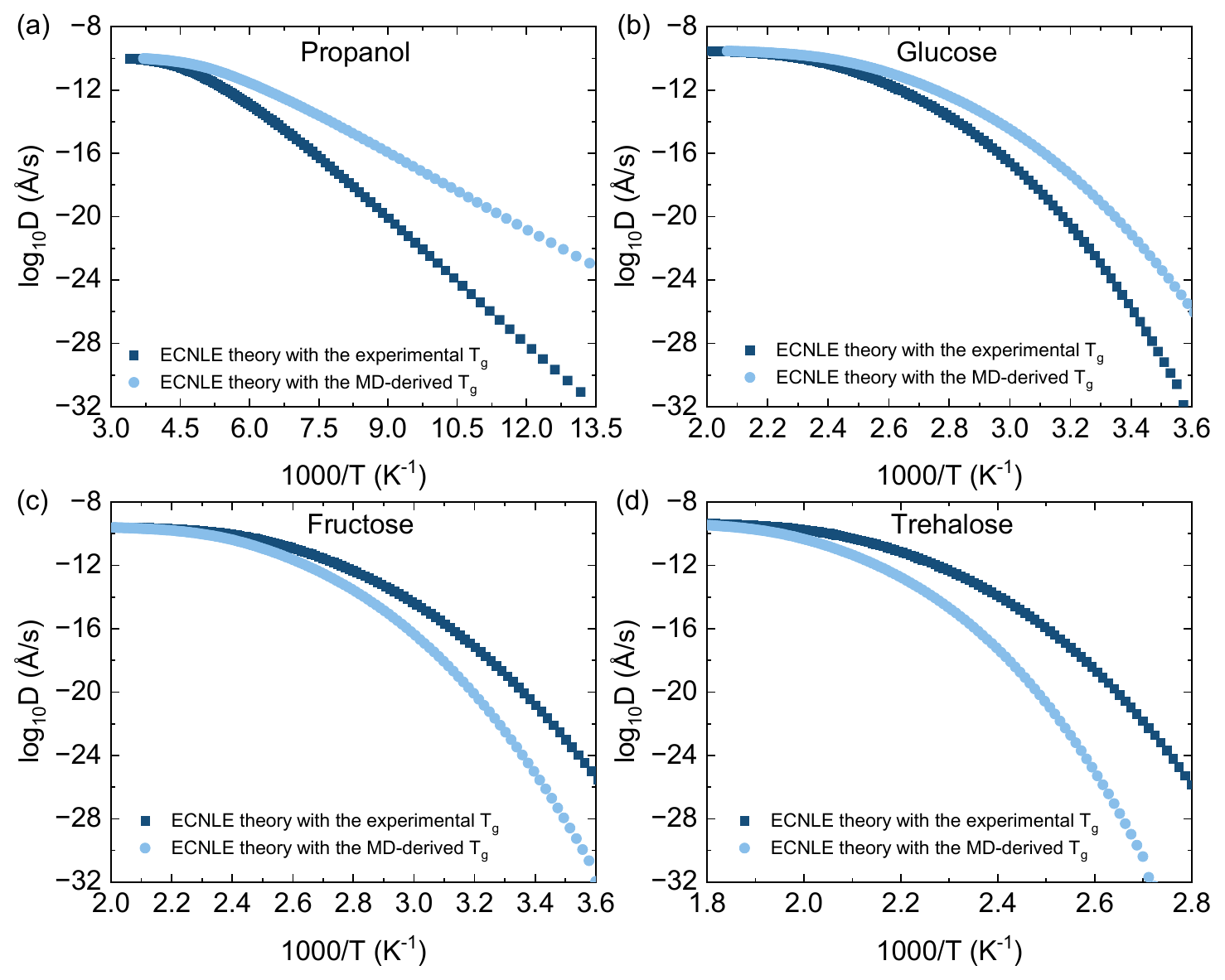}
\caption{\label{fig5}(Color online) {Logarithm of the diffusion coefficient as a function of $1000/T$ for (a) propanol, (b) glucose, (c) fructose, and (d) trehalose. Square symbols represent ECNLE theory predictions using the experimental $T_g$, while circle symbols correspond to ECNLE predictions based on our simulation $T_g$ values.}}
\end{figure*}

Following the relaxation time analysis above, Fig. \ref{fig5} shows theoretical predictions and experimental data of the diffusion coefficient for propanol, glucose, fructose, and trehalose. Again, we present two sets of diffusion coefficient predictions for each material: one using the experimental $T_g$ and the other using MD-predicted $T_g$ values. In all cases, the diffusion coefficients exhibit strong non-Arrhenius behavior as the systems approach the glass transition. The diffusion trends predicted using both $T_g$ inputs agree well at higher temperatures, while noticeable deviations at lower temperatures suggest an increasing influence of cooperative dynamics. These are predictive results that can be further tested and validated by future molecular simulations and experimental measurements.

\section{Conclusion}
In conclusion, we have presented an integrated approach that combines MD simulations with the ECNLE theory to investigate the relaxation dynamics of small organic molecules, namely propanol, glucose, fructose, and trehalose. The glass transition temperatures were obtained from the simulations and used as key input to the ECNLE model to predict the temperature dependence of structural relaxation times and diffusion coefficients. The resulting theoretical predictions show excellent agreement with the experimental data and capture essential features such as non-Arrhenius behavior and the pronounced dynamic slowdown near $T_g$. This framework extends the accessible timescales far beyond the limits of conventional MD simulations and does so without any empirical fitting. Good predictive performance across multiple systems clearly indicates the generality and transferability of our approach.
\begin{acknowledgments}
This research was funded by the Vietnam National Foundation for Science and Technology Development (NAFOSTED) under Grant No.103.01-2023.62.
\end{acknowledgments}
\section*{Supplementary material}
Supplementary material associated with this article can be found in the online version.
\section*{Conflicts of interest}
The authors have no conflicts to disclose.

\section*{Data availability}
The data that support the findings of this study are available from the corresponding author upon reasonable request.

\end{document}